\documentclass[%
 reprint,
superscriptaddress,
 amsmath,amssymb,
 aps,
]{revtex4-2}

\usepackage{graphicx                                }
\usepackage{dcolumn}
\usepackage{bm}
\usepackage{hyperref}
\hypersetup{
    colorlinks,
    linkcolor={red},
    citecolor={blue},
    urlcolor={blue}
}

\newcommand{\iu}{{i\mkern1mu}}

\DeclareMathOperator{\sign}{sign}

\begin{document}

\preprint{APS/123-QED}

\title{Universal momentum tail of identical one-dimensional anyons with two-body interactions}

\author{Ra\'ul Hidalgo-Sacoto}
\email{raul.hidalgo@oist.jp}
\affiliation{%
 Quantum Systems Unit, Okinawa Institute of Science and Technology Graduate University, \\ Onna, Okinawa, Japan}%
 
\author{Thomas Busch}%
 \email{thomas.busch@oist.jp}
\affiliation{%
 Quantum Systems Unit, Okinawa Institute of Science and Technology Graduate University, \\ Onna, Okinawa, Japan}%

\author{D. Blume}
\email{doerte.blume-1@ou.edu}
\affiliation{Homer L. Dodge Department of Physics and Astronomy,
  The University of Oklahoma,
  440 W. Brooks Street,
  Norman,
Oklahoma 73019, USA}
\affiliation{Center for Quantum Research and Technology,
  The University of Oklahoma,
  440 W. Brooks Street,
  Norman,
Oklahoma 73019, USA}

\date{\today}

\begin{abstract}
Non-relativistic anyons in 1D possess generalized exchange statistics in which the exchange of two identical anyons generates a non-local phase that is governed by the spatial ordering of the particles and the statistical parameter $\alpha$. Working in the continuum, we demonstrate the existence of two distinct types of 1D anyons, namely bosonic anyons and fermionic anyons. We identify a many-body Hamiltonian with additive two-body zero-range interactions that supports bosonic and fermionic anyon eigenstates, which are, for arbitrary interaction strength, related through a generalized bosonic-anyon—fermionic-anyon mapping, an extension of the celebrated Bose-Fermi mapping for zero-range interacting 1D systems. The momentum distributions of bosonic and fermionic anyons are distinct: while both feature $k^{-2}$ and $k^{-3}$ tails, the associated prefactors differ. Our work reveals intricate connections between the generalized exchange statistics, the universal two- and three-body Tan contacts of systems consisting of $N$ identical particles, and the emergence of statistics-induced chiral symmetry breaking.
\end{abstract}

\maketitle

{\em{Introduction:---}}
The exchange symmetries of a quantum system profoundly impact its properties~\cite{leinaas1977theory}.
 Upon exchange of two identical elementary particles, the many-body wave function remains unchanged for two identical bosons
 and picks up a minus sign for two identical fermions~\cite{baym2018lectures}. The interplay of exchange statistics and particle interactions 
 gives rise to a myriad of intriguing phenomena, including many-body quantum phase transitions~\cite{bloch2008many} or the emergence
 of exotic few-particle Efimov states~\cite{naidon2017efimov}. In this work we focus on 1D single-component quantum systems, which can be realized experimentally by confining 
 ultracold atoms into a waveguide or an optical lattice~\cite{cazalilla2011one,mistakidis2023fewbody}.
The interactions in such systems can be tuned from weak to strong (attractive or repulsive) by either varying the transverse confinement or the scattering length in the vicinity of a Feshbach resonance~\cite{olshanii1998atomic, bergeman2003atom, granger2004tuning, chin2010feshbach}. In general, quantum correlations are enhanced in low-dimensional systems~\cite{mermin1966absence, haldane1981effective, petrov2000regimes, giamarchi2003quantum}. A unique aspect of 1D systems is that the particles 
 can be ordered and that they need to “pass through each other” to exchange positions~\cite{girardeau1960relationship}. Correspondingly, particle exchange is intimately linked to particle interactions.
 
 It is well established that strongly-interacting bosons (Tonks-Girardeau bosons) behave, in many respects, like non-interacting fermions~\cite{girardeau1960relationship}.  Intriguingly, the correspondence exists for any interaction strength or scattering length: the Bose-Fermi mapping transforms the wave function of $N$ identical bosons that interact via a sum of two-body even-parity zero-range interactions  with 1D scattering length $a_+$ to the wave function of $N$ identical fermions that interact via a sum of two-body odd-parity zero-range interactions 
 with 1D scattering length $a_-$, provided $a_+$ and $a_-$ are equal to each other~\cite{cheon1998realizing, cheon1999fermion, girardeau2004effective, girardeau2003fermi}. 
Signatures of the Bose-Fermi duality have, e.g., been probed experimentally at the two-atom level via tunneling spectroscopy~\cite{zurn2012fermionization}.
Importantly,  non-local observables such as the momentum distribution unambiguously reveal that bosons and fermions are fundamentally distinct objects, even though their 
 many-body wave functions are intimately connected~\cite{lenard1964momentum, bender2005exponentially}.
Going beyond “traditional” bosonic and fermionic exchange statistics, anyons obey generalized fractional statistics~\cite{leinaas1977theory, haldane1991fractional}. While studied primarily in the context of two-dimensional systems~\cite{wilczek1990fractional, stern2008anyons}, the flexibility afforded by ultracold atom systems has motivated experimental and theoretical efforts on lattice anyons in varying dimensions~\cite{keilmann2011statistically, greschner2015anyon, munoz2020anyonic, bonkhoff2021bosonic, leonard2023realization, kwan2024realization}.

In this work we consider $N$ identical anyons in 1D, for which the many-body wave function picks up a chiral fractional phase upon exchange of two identical anyons~\cite{kundu1999exact,girardeau2006anyon,batchelor2006one, pactu2007correlation, hao2008ground, del2008fermionization, piroli2017exact, wang2024boson}. Identical anyons in the continuum were first introduced by Kundu by applying a continuous Jordan-Wigner transformation to bosonic field operators~\cite{kundu1999exact} and later by Girardeau by applying a continuous Jordan-Wigner transformation to fermionic field operators~\cite{girardeau2006anyon}. Very recently~\cite{wang2024boson}, identical anyons with contact interactions were constructed from a two-component Fermi system with $a_+=a_-$ by forming linear combinations of spin-triplet and spin-singlet dimer wave functions, yielding an anyonic superfluid. Wang {\em{et al.}}~\cite{wang2024boson} identified the anyonic scattering length $a_{\text{any}}$ as the fundamental microscopic two-body parameter and established the existence of a boson-anyon-fermion mapping.  
A few months ago, 1D anyons in the continuum were realized  for the first time experimentally by  utilizing the low-energy spin excitations of a Bose gas with a single  impurity~\cite{dhar2024anyonization}.

\begin{figure}[tb]
    \centering
\includegraphics[width=0.9\linewidth]
    {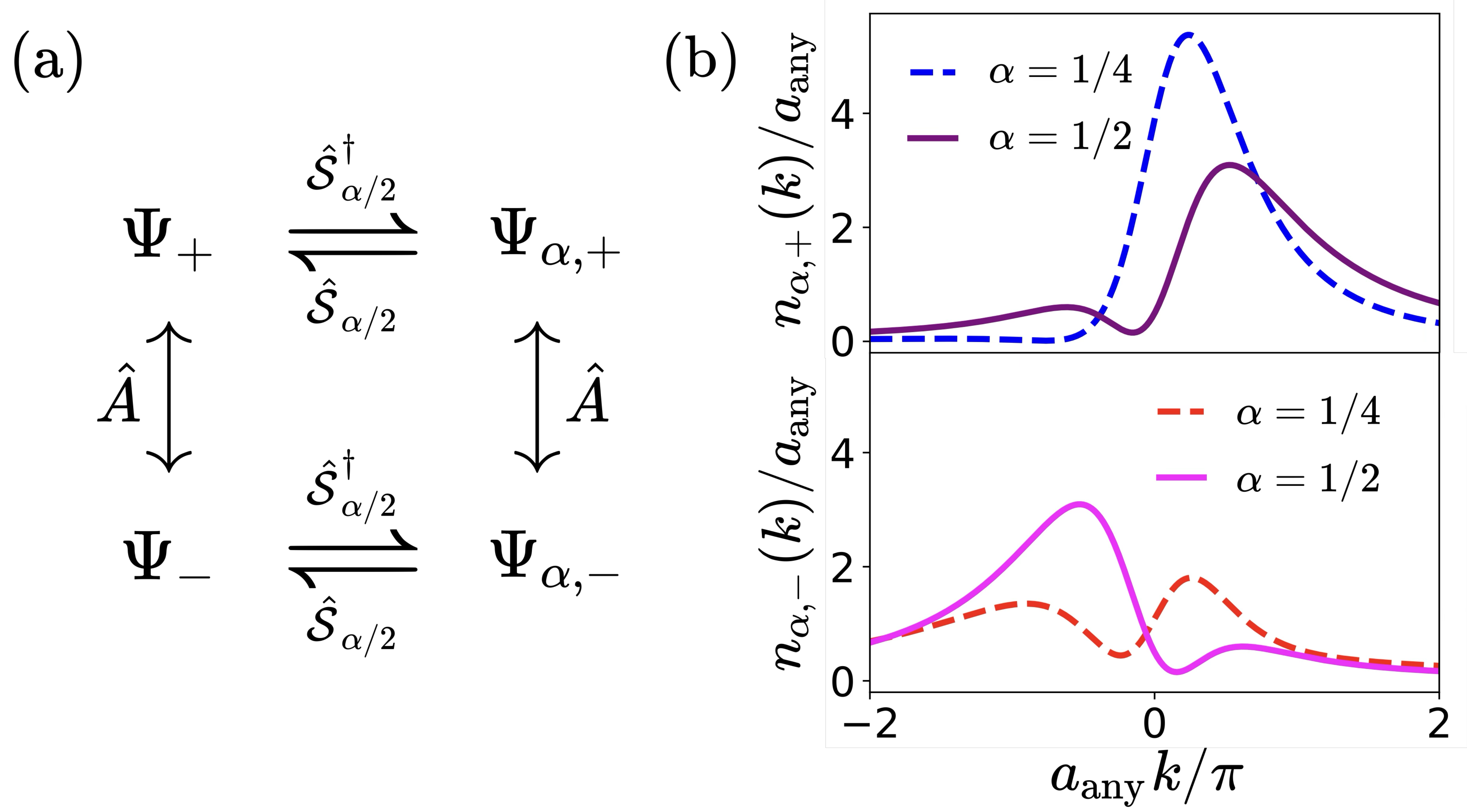}
    \caption{(a)
    Bosonic-anyon --- fermionic-anyon mapping that connects $\Psi_+$, $\Psi_-$, $\Psi_{\alpha,+}$, and $\Psi_{\alpha,-}$. 
    (b) Momentum 
    distributions (top) $n_{\alpha,+}(k)$ and (bottom) $n_{\alpha,-}(k)$ for three identical free-space bosonic anyons and  fermionic anyons, respectively,  with $\alpha=1/4$ (dashed lines) and $\alpha=1/2$ (solid lines).  The momentum distributions $n_{\alpha,\pm}(k)$  are asymmetric with respect to $k=0$ due to  the hybridization of the symmetric and anti-symmetric parts of the anyonic wavefunction.}
    \label{fig:anyon-map}
\end{figure}

The key results of this Letter are: (A) We demonstrate that wave functions that obey anyonic exchange statistics can be categorized into two distinct types: bosonic anyons or fermionic anyons. (B) Inspired by a composite particle picture~\cite{fradkin2017disorder, valenti2023topological}, we construct bosonic and fermionic anyon wavefunctions by  “dressing”, respectively, bosonic and fermionic wavefunctions with a statistical (i.e., $\alpha$-dependent) gauge phase. If the initial bosonic and fermionic wave functions themselves are related via the Bose-Fermi mapping~\cite{girardeau2003fermi}, then the wavefunctions of bosonic anyons and those of fermionic anyons are related via a bosonic-anyon—fermionic-anyon mapping [see Fig.~\ref{fig:anyon-map}(a)]. Findings (A) and (B) allow one to reconcile differing existing definitions of 1D anyons~\cite{kundu1999exact, girardeau2006anyon, calabrese2007correlation, santachiara2008one, zinner2015strongly,del2008fermionization}. (C) We introduce a Hamiltonian with additive even- and odd-parity zero-range interactions that supports both bosonic and fermionic anyon eigenstates. (D) We show that bosonic anyons and fermionic anyons are—much like bosons and fermions—distinct particles that have unique momentum distributions [see Fig.~\ref{fig:anyon-map}(b)]. Specifically, we find that the sub-leading chiral $k^{-3}$  term of the momentum distributions of bosonic anyons and fermionic anyons differs.

{(A)\em{\:Anyonic exchange statistics ---}} We consider $N$ identical  non-relativistic spinless quantum particles of mass $m$ in 1D with position coordinates $z_1,\dots,z_N$.
If $\Psi(z_1,\dots,z_N)$ with 
normalization
$\int dz_1 \cdots dz_N |\Psi(z_1, \dots,z_N)|^2 = 1$
denotes an un-symmetrized $N$-particle wave function, then
a fully symmetric bosonic wave function $\Psi_+$ or fully anti-symmetric fermionic wave function 
$\Psi_-$ can be constructed
via~\cite{baym2018lectures}
\begin{equation}
    \label{eq_general_symm}
    \Psi_{\pm}(z_1,\dots,z_N)=\frac{1}{\sqrt{N!}}
    \sum_{\text{perm}} (\pm1)^P \hat{\cal{P}} \Psi(z_1,\dots,z_N),
\end{equation}
where the sum runs over all $N!$ particle orderings that are generated by 
the permutation operator $\hat{\cal{P}}$ and $P$ is the number of two-particle exchanges required to realize a given particle ordering.
Under the exchange of particles $k$ and $j$, i.e., under application of the exchange operator $\hat{P}_{kj}$,
the wave functions $\Psi_\pm$ obey the usual exchange symmetry 
conditions
\begin{align}
    \hat{P}_{kj}\Psi_{\pm}(z_1, &\dots,z_k,\dots,z_j, \dots, z_N) = \nonumber \\
    &\pm \Psi_\pm(z_1, \dots,z_j,\dots,z_k, \dots, z_N).
    \label{eq:particleexchange}
\end{align}

We define the anyonic exchange symmetry by demanding that the many-body $N$-anyon wave function $\Psi_{\alpha,\pm}$, where the statistical parameter $\alpha$ ranges from $0$ to $1$, picks up a non-local 
phase upon exchange of anyons $k$ and $j$,
\begin{align}
\label{eq:anyonicexchangeN}
    \hat{P}_{kj} &\Psi_{\alpha,\pm}(z_1,\dots,z_k,\dots,z_j,\dots, z_N) = \nonumber \\
    &\pm e^{-\iu \pi \alpha \left[ \sum_{l=j+1}^{k-1}\sign(z_{lk}) + \sum^{k}_{l=j+1}\sign(z_{jl})\right]} \times
    \nonumber  \\
    &\qquad\qquad\Psi_{\alpha,\pm}(z_1,\dots,z_j,\dots,z_k,\dots, z_N),
\end{align}
where $j<k$ and $z_{lk}=z_l-z_k$. The sign-function $\sign(z)$ is defined to be equal to $-1$ and $+1$ for $z<0$ and $z>0$, respectively, and undefined for $z=0$~\cite{SM, wang2024boson}; moreover, $\sign^2(z)=1$ for all $z$ ~\cite{ValienteBFdualities}.
The jump discontinuities of the $\sign$-functions contained in the non-local phase have, as will be discussed below, important physical implications. 
We emphasize that the  exchange condition (\ref{eq:anyonicexchangeN}) defines  two types of anyons, namely bosonic anyons described by $\Psi_{\alpha,+}$, which have an overall plus sign on the right hand side of Eq.~(\ref{eq:anyonicexchangeN}), and fermionic anyons described by $\Psi_{\alpha,-}$, which have an overall minus sign on the right hand side of Eq.~(\ref{eq:anyonicexchangeN}).
The literature either considers bosonic or fermionic anyons~\cite{kundu1999exact, girardeau2006anyon, calabrese2007correlation, santachiara2008one, zinner2015strongly}, 
including anyonic Tonks-Girardeau bosons and
anyonic Tonks-Girardeau fermions~\cite{del2008fermionization}. 
We will show below that our construction of $\Psi_{\alpha,\pm}$ reveals a deep connection between $\Psi_{\alpha,+}$ and $\Psi_{\alpha,-}$ while the tail of the momentum distributions
$n_{\alpha,\pm}(k)$ reveals their distinctness.

{(B)\em{\:Construction of anyonic wave functions $\Psi_{\alpha,\pm}$ and bosonic-anyon---fermionic-anyon mapping ---}}
Our construction of $N$-anyon wave functions is inspired by a composite-particle-picture~\cite{valenti2023topological}, which provides a general framework for constructing and interpreting quasi-particles. We consider a complete set of fully symmetrized functions $\Psi_{+}$ and a complete set of fully anti-symmetrized functions $\Psi_{-}$, which are chosen such that they are related via the well-established Bose-Fermi mapping~\cite{girardeau2003fermi},
\begin{eqnarray}
\label{eq_bf_mapping}
\Psi_{\pm}(z_1,\dots,z_N)= 
\hat{A}(z_1,\dots,z_N)\Psi_{\mp}(z_1,\dots,z_N),
\end{eqnarray}
where 
\begin{eqnarray}
\label{eq_BF_mapping_capA}
\hat{A}(z_1,\dots,z_N)=\prod_{j=1}^{N-1} \prod_{k=j+1}^N \sign(z_{jk}).
\end{eqnarray} 
We define the operator $\hat{{\cal{S}}}_{\alpha/2}^{\dagger}(z_1,\cdots,z_N)$,
\begin{eqnarray}
\hat{{\cal{S}}}_{\alpha/2}^{\dagger}(z_1,\dots,z_N)=
\left( \prod_{j=1}^{N-1} \prod_{k=j+1}^N {\cal{N}}_{\alpha} \hat{S}_{\alpha/2}^{\dagger}(z_{jk}) 
\right),
\end{eqnarray}
where the phase factor or normalization constant ${\cal{N}}_{\alpha}$ ($|{\cal{N}}_{\alpha}|^2=1$) is given by 
\begin{eqnarray}
    {\cal{N}}_{\alpha} = \left[ (1-\alpha)^2+\alpha^2  \right]^{-1/2} \left[
    (1-\alpha) -\iu \alpha \right],
\end{eqnarray}
in terms of the anyonic two-particle exchange operator 
$\hat{S}_{\alpha}(z_{jk})$,
\begin{eqnarray}
    \hat{S}_{\alpha}(z_{jk}) = e^{-\iu \pi \alpha \sign(z_{jk})}.
\end{eqnarray}
The operator $\hat{\cal{S}}_{\alpha/2}^{\dagger}$
can be interpreted, motivated by ideas introduced in the context of field theories~\cite{fradkin1978order}  and  characterization schemes of global quantum phases via singular local operators~\cite{chen2022topological, fradkin2017disorder}, as a disorder operator~\cite{kadanoff1969operator, kadanoff1971determination}.
 With the above definitions, an explicit calculation shows that the $\Psi_{\alpha,\pm}$, 
 constructed via
\begin{eqnarray}
\label{eq_anyon_contruction}
    \Psi_{\alpha,\pm}(z_1,\dots,z_N)=
    \hat{{\cal{S}}}_{\alpha/2}^{\dagger}(z_1,\dots,z_N)
\Psi_{\pm}(z_1,\dots,z_N),\nonumber \\
\end{eqnarray}
obey anyonic exchange statistics.
The phase factor  ${\cal{N}}_{\alpha}$
is chosen such that for $\alpha=0$ the $\Psi_{\alpha,\pm}$ coincide with the corresponding “native” wave functions,
\begin{eqnarray}
\Psi_{\alpha=0,\pm}(z_1,\cdots,z_N)=\Psi_{\pm}(z_1,\cdots,z_N),
\end{eqnarray}
and with the dual complements of the  “native” wave functions --- defined via Eq.~(\ref{eq_bf_mapping}) --- for $\alpha=1$,
\begin{eqnarray}
\label{eq_norm2}
\Psi_{\alpha=1,\pm}(z_1,\cdots,z_N)= \quad \quad \quad \quad \quad \quad \quad \\ \nonumber
\hat{A}(z_1,\cdots,z_N)\Psi_{\pm}(z_1,\cdots,z_N).
\end{eqnarray}
Equation~(\ref{eq_anyon_contruction}) states that anyonic wave functions can be generated by 
imprinting 
a gauge 
phase
onto
wave functions that describe naturally occurring particles, namely collections of identical bosons or collections of identical fermions. This not only suggests that bosonic (fermionic) anyons can be interpreted as being generated by “dressing” bosons (fermions) but also identifies a starting point for designing protocols  for the experimental realization of anyons.
Using that $\Psi_{+}$ and $\Psi_-$
are orthogonal to each other, it can be checked readily that $\Psi_{\alpha,+}$ and $\Psi_{\alpha,-}$
are also orthogonal to each other; this observation indicates that bosonic and fermionic anyons are different quasi-particles that can be distinguished by their momentum distributions (see below).

Equations~(\ref{eq_bf_mapping}) and (\ref{eq_anyon_contruction}) imply the bosonic-anyon---fermionic-anyon mapping schematically depicted in 
Fig.~\ref{fig:anyon-map}(a).
In identifying the mapping, we used that $\hat{A}$ is its own inverse
and that $(\hat{\cal{S}}_{\alpha/2}^\dagger)^{-1}$  is equal to
$\hat{\cal{S}}_{\alpha/2}$.
Provided the Bose-Fermi mapping
[left vertical arrow in Fig.~\ref{fig:anyon-map}(a)]
holds,
bosonic and fermionic anyons can be constructed by applying the transformations indicated on the arrows, either
following the “loop” clock-wise or anti-clock-wise.  
We emphasize that the anyonic exchange operator critically hinges on the fact that the particles are constrained to move in one spatial dimension; the reason is that the ordering of particles is only well defined in one spatial dimension. This is the same reason for why the Bose-Fermi mapping is specific to 1D systems~\cite{cheon1999fermion, girardeau2003fermi}.

{(C)\em{\:Hamiltonian for anyons with two-body zero-range interactions ---}}
We next identify a many-body Hamiltonian $\hat{H}$ with two-body zero-range interactions that supports eigenstates that possess bosonic anyon or fermionic anyon statistics.
To start with, we recall that $\Psi_{+}$ and $\Psi_-$ are solutions to the Schr\"odinger equation for the Hamiltonians $\hat{H}_+$ and $\hat{H}_-$, 
where $\hat{H}_+=\hat{T}+ 
\sum_{j<k}\hat{V}_+(z_{jk})$,
$\hat{H}_-=\hat{T}+ 
\sum_{j<k}\hat{V}_-(z_{jk})$,
and
$\hat{T}$ denotes the $N$-particle kinetic energy operator~\cite{lieb1963exact}. The two-body zero-range pseudo-potentials $\hat{V}_{\pm}(z_{jk})$ read 
\begin{align}
    \hat{V}_+(z_{jk})&=g_+\delta(z_{jk})\\
    \hat{V}_-(z_{jk})&=
    g_-\frac{\overleftarrow{\partial}}{\partial z_{jk}}\delta(z_{jk})
    \frac{\overrightarrow{\partial}}{\partial z_{jk}}.
\end{align}
For $g_+g_-=-4\hbar^4/m^2$ or $a_+=a_-$ [$g_+=-2 \hbar^2/(m a_+)$
and $g_-=2 \hbar^2 a_-/m$], the eigenstates $\Psi_+$ and $\Psi_-$ are related via
Eq.~(\ref{eq_bf_mapping})~\cite{girardeau2003fermi}.
Importantly, the two-body zero-range interactions can alternatively be formulated
by specifying the  behavior of $\Psi_{\pm}$ in the limit that the interparticle distance $z_{jk}$ for any two particles $j$ and $k$ goes to zero while all other coordinates [including the two-body center-of-mass coordinate $Z_{jk}$, $Z_{jk}=(z_j+z_k)/2$] are held fixed~\cite{olshanii2003short},
\begin{eqnarray}
    \Psi_{\pm}(z_1,\dots,z_N) \underset{z_{jk}\rightarrow 0}{\longrightarrow}
    h_{\pm}^{(2)}(z_{jk})
    \Phi_{\pm}(Z_{jk},\{z_l\}_{l \ne j,k}).
\end{eqnarray}
Here, $\Phi_{\pm}$ is a $z_{jk}$-independent “amplitude” while $h_{\pm}^{(2)}(z_{jk})$ describes the short-distance behavior (up to sub-leading order in $|z_{jk}|$) imposed by the zero-range interactions $\hat{V}_{\pm}(z_{jk})$:  
$h_+^{(2)}(z_{jk})=1-|z_{jk}|/a_+$
and $h_-^{(2)}(z_{jk})=\sign(z_{jk})-z_{jk}/a_{-}$.
Since the leading term of $h_+^{(2)}(z_{jk})$ is a constant, adding the potential $\sum_{j<k}\hat{V}_-(z_{jk})$ to $\hat{H}_+$ does not change its eigenenergies and eigenstates.
Similarly, since the leading term of $h_-^{(2)}(z_{jk})$ is equal to $\sign(z_{jk})$, adding the potential $\sum_{j<k}\hat{V}_+(z_{jk})$ to $\hat{H}_-$ does not change its eigenenergies and eigenstates.
As a consequence,  $\Psi_{+}$ and $\Psi_-$ are  both eigenstates of $\hat{H}$,
\begin{eqnarray}
    \label{eq_ham_anyon}
    \hat{H}=\hat{T}+\sum_{j<k} \hat{V}_+(z_{jk})+
    \sum_{j<k} \hat{V}_-(z_{jk}),
\end{eqnarray}
provided the coupling strengths $g_{\pm}$ are chosen such that $a_+=a_-$; importantly, this condition guarantees $h_-^{(2)}(z_{jk})=\sign(z_{jk})h_+^{(2)}(z_{jk})$.

It can now be readily shown that the anyonic states $\Psi_{\alpha,\pm}$, defined via Eq.~(\ref{eq_anyon_contruction}), obey the two-particle boundary conditions implied by Eq.~(\ref{eq_ham_anyon}). To see this,
we set $a_+=a_-=a_{\text{any}}$ and look at
$\Psi_{\alpha,\pm}$ in the $z_{jk} \rightarrow 0$ limit. Rewriting $\hat{S}^\dagger_{\alpha/2}(z_{jk}) = \cos (\pi \alpha/2) + \iu \sin (\pi \alpha/2)\sign(z_{jk})$, the $z_{jk}$-dependent term in $\Psi_{\alpha,\pm}$, i.e., the term  
\begin{align}
    \label{eq_limit_twobody}
    \hat{S}_{\alpha/2}^{\dagger}&(z_{jk}) h_{\pm}^{(2)}(z_{jk})
    \underset{z_{jk}\rightarrow 0}{\longrightarrow}
    \nonumber \\
    &\cos(\pi \alpha/2) h_{\pm}^{(2)}(z_{jk})
    + \iu \sin(\pi \alpha/2) h_{\mp}^{(2)}(z_{jk}),
\end{align}
shows that the anyonic two-body exchange operator creates a superposition of
the even- and odd-parity short-distance behaviors; both of these terms obey the boundary conditions implied by  $\hat{H}$, Eq.~(\ref{eq_ham_anyon}). 
Using the same logic as was used when deriving the Bose-Fermi mapping~\cite{girardeau2003fermi}, 
it then follows that the $\Psi_{\alpha,\pm}$ are eigenstates of $\hat{H}$.

{(D)\em{\:Momentum tail of $N$ identical bosonic and fermionic anyons ---}}
 Using the anyonic eigenstates $\Psi_{\alpha,\pm}$  of $\hat{H}$ [Eqs.~(\ref{eq_anyon_contruction}) and (\ref{eq_ham_anyon})], we now derive analytical expressions
for the large-$|k|$ behavior of the momentum distributions 
$n_{\alpha,\pm}(k) $,
which are defined
through~\cite{sekino2018comparative}
\begin{align}
    n_{\alpha,\pm}(k) = N\int &dz_2 \dots dz_N 
    \times 
    \nonumber \\
    & 
    \left| \int dz_1 \: e^{-\iu k z_1} \Psi_{\alpha,\pm}(z_1,...,z_N)\right|^2 .\label{eq:momentum_correlationfunction1}
\end{align}
We will show that the momentum distribution tails, up to order $k^{-3}$, can be expressed in terms of the statistical parameter $\alpha$, the anyonic scattering length $a_{\text{any}}$, and the universal two- and three-body Tan contacts ${\cal{C}}_2$ and ${\cal{C}}_3$ of the $N$-particle system.
To determine the large-$|k|$ tails of $n_{\alpha,\pm}(k)$, we utilize two properties of Fourier transforms~\cite{bleistein1986asymptotic}. (i) A frequency spectrum only contains an algebraic tail at large frequencies if the time signal, or a (higher-order) derivative thereof, contains discontinuities. (ii) Discontinuities that are separated in time contribute independently to the algebraic fall-off of the frequency spectrum.

Applying these properties to the position-momentum domains, 
it was shown previously that the slowest decay of the 1D momentum tails of $N$ identical bosons and $N$ identical fermions with zero-range interactions is given by $4 {\cal{C}}_2 (a_+)^{-2} k^{-4} $~\cite{olshanii2003short, sekino2018comparative} and $4 {\cal{C}}_2 k^{-2}$ ~\cite{cui2016universal,sekino2018comparative}, respectively. 
The discontinuities that give rise to these tails are the $-|z_{jk}|/a_+$ piece of $h_+^{(2)}(z_{jk})$ for bosons 
and the $\sign(z_{jk})$ piece of $h_-^{(2)}(z_{jk})$ for fermions. 

For $N$ identical anyons, discontinuities of the eigenstates and their derivatives are not only arising from the two-body zero-range boundary conditions that are encoded in  $h_{\pm}^{(2)}(z_{jk})$ but also 
 from the statistical gauge  operator $\hat{{S}}_{\alpha/2}^{\dagger}(z_{jk})$. Even though $\hat{H}$ is fully defined without an explicit three-body potential, the dressing of $\Psi_{\pm}$ by the gauge 
 phase 
 introduces three-body correlations into the momentum tail at order $k^{-3}$
 (for details, see~\cite{SM}).
Motivated by these considerations, we consider two types of discontinuities: (1) two-body discontinuities that arise when $z_q$ is equal to $z_1$ but $z_l$ ($l=2,\dots,q-1,q+1,\dots,N$) is not equal to  $z_1$ or $z_q$; and (2) three-body discontinuities that arise when $z_j$ and $z_q$ are equal to $z_1$ but $z_l$ ($l=2,\dots,j-1,j+1,\dots,q-1,q+1,\dots,N$) is not equal to  $z_1$, $z_j$, or $z_q$. There exist $N-1$ 
allowed $q$ values that need to be considered when analyzing the two-body discontinuities and $(N-1)(N-2)/2$ allowed $j$ and $q$ combinations that need to be considered when analyzing the three-body discontinuities. 
Because each  discontinuity contributes additively to the tail of the momentum distribution, the large-$|k|$ momentum behavior can be written as (up to order $k^{-3}$)
\begin{align}
     \lim_{|k| \rightarrow \infty}& n_{\alpha,\pm}(k) = \nonumber \\ &(N-1) T^{(2)}_{\alpha,\pm}(k) + \frac{(N-1)(N-2)}{2}T^{(3)}_{\alpha,\pm}(k),
\end{align}
     where
     \begin{align}
     T^{(2)}_{\alpha,\pm}(k)&=
      \lim_{ |k| \rightarrow \infty,z_1 \rightarrow z_2}   n_{\alpha,\pm}(k), \\
       T^{(3)}_{\alpha,\pm}(k)&=
      \lim_{|k| \rightarrow \infty, z_2 \rightarrow z_3, z_1 \rightarrow Z_{23} }n_{\alpha,\pm}(k) .
   \end{align}
The quantity $T^{(2)}_{\alpha, \pm}$  depends on the limiting two-body behavior of the wavefunction $\Psi_{\alpha,\pm}$, i.e., on
$h^{(2)}_{\alpha,\pm}(z_{12}) =  \hat{S}^\dagger_{\alpha/2}(z_{12}) h^{(2)}_{\pm}(z_{12})$ [see Eq.~(\ref{eq_limit_twobody})].
The quantity $T^{(3)}_{\alpha, \pm}$, in contrast, depends on the limiting three-body behavior of the wavefunction $\Psi_{\alpha,\pm}$, i.e., on
$h^{(3)}_{\alpha,\pm}(z_{12},z_{12,3}) =  \hat{\cal{S}}^\dagger_{\alpha/2}(z_{12},z_{13},z_{23}) h^{(3)}_{\pm}(z_{12},z_{12,3})$,
where $z_{12,3}=(z_1+z_2)/2-z_3$.
Using the bosonic exchange symmetry, one can show that the leading-order behavior of $h^{(3)}_{+}(z_{12},z_{12,3})$ is given by $1$.

These limiting two- and three-body behaviors determine the momentum tails of bosonic anyons and fermionic anyons up to 
order 
$k^{-3}$, 
\begin{align}
    \label{eq_momentumtail_alpha+}
       \lim_{|k| \rightarrow \infty} n_{\alpha,+}(k) =&
       \frac{4  {\cal{C}}_2}{k^2}  \sin^2 \left( \frac{\pi \alpha}{2} \right) + 
       \frac{4  {\cal{C}}_2}{a_{\mathrm{any}}k^3}
         \sin \left( \pi \alpha \right)
        \nonumber \\
       &+\frac{8 {\cal{C}}_3}{k^3} \sin ^2\left( \frac{\pi \alpha}{2} \right) \sin(\pi \alpha)  
      +{\cal{O}}(k^{-4}) \quad
\end{align}
and
\begin{align}
    \label{eq_momentumtail_alpha-}
       \lim_{|k| \rightarrow \infty} n_{\alpha,-}(k) = &
       \frac{4  {\cal{C}}_2}{k^2}  \cos^2 \left( \frac{\pi \alpha}{2} \right) - 
       \frac{4  {\cal{C}}_2}{a_{\mathrm{any}}k^3}
         \sin \left( \pi \alpha \right)
        \nonumber \\
       &-\frac{8 {\cal{C}}_3}{k^3} \cos ^2\left( \frac{\pi \alpha}{2} \right) \sin(\pi \alpha)  
      +{\cal{O}}(k^{-4}). \quad
\end{align}
The next-order term, i.e., the $k^{-4}$ contribution, requires knowledge of the sub-sub-leading short-distance two-body behavior, the sub-leading short-distance three-body behavior, and the leading short-distance four-body behavior.
The Tan contacts ${\cal{C}}_2$ and ${\cal{C}}_3$ \cite{tan2008energetics, tan2008large, barth2011tan}, which enter into $n_{\alpha,\pm}(k)$, are  
defined in terms of the two- and three-particle   probability density distributions 
$g_2(z,z)$ and 
$g_3(z,z,z)$~\cite{sekino2018comparative};
specifically, 
$\mathcal{C}_2 =\int dz g_2(z,z)$ and $\mathcal{C}_3 =\int dz g_3(z,z,z)$,
where 
\begin{align}
\label{eq_density_capm}
    g_M(z_1,& \dots, z_M) =\nonumber\\
    &\frac{N!}{(N-M)!}\int dz_{M+1}\dots dz_{N} 
    \left|\Psi_{\alpha,\pm}(z_1, \dots, z_N)\right|^2. \quad
\end{align}
Importantly, for 1D systems with two-body zero-range interactions characterized by $a_+=a_-=a_{\text{any}}$, ${\cal{C}}_2$ and ${\cal{C}}_3$ are independent of the particle statistics, i.e., the Tan contacts that appear in Eqs.~(\ref{eq_momentumtail_alpha+}) and (\ref{eq_momentumtail_alpha-}) are the same for $N$ identical bosons, $N$ identical fermions,  $N$ identical bosonic anyons, and $N$ identical fermionic anyons.
The universal relations, Eqs.~(\ref{eq_momentumtail_alpha+}) and (\ref{eq_momentumtail_alpha-}), hold for any number of particles, interaction strength,
or trapping potential. 
We have confirmed our predictions for the  momentum tails through explicit calculations of the momentum distribution for two and three identical anyons with zero-range interactions in free space~\cite{SM}.

Assuming fixed $\alpha$ and  $a_{\text{any}}$,
Eqs.~(\ref{eq_momentumtail_alpha+}) and (\ref{eq_momentumtail_alpha-}) show that the momentum tails for bosonic anyons and fermionic anyons differ.
For $\alpha=0$, the $k^{-2}$ and $k^{-3}$ terms of $n_{\alpha,+}(k)$ vanish; this is consistent with the fact that the leading-order term of $N$ identical bosons is 
$k^{-4}$~\cite{olshanii2003short}.
In the same limit,  the $k^{-2}$ and $k^{-3}$ terms of $n_{\alpha,-}(k)$ go to
$4{\cal{C}}_2 k^{-2}$ and $0$, which is consistent with the established behavior for identical fermions~\cite{cui2016universal}. 
For $\alpha=1$, in contrast, the asymptotic behaviors of $n_{\alpha,+}(k)$ and $n_{\alpha,-}(k)$ are --- as expected based on Eq.~(\ref{eq_norm2}) --- reversed.
 The $k^{-3}$ term only exists for $0<\alpha<1$ and finite $a_{\text{any}}$ ($a_{\text{any}}\ne 0$ and $a_{\text{any}} \ne \pm \infty$).
 A $k^{-3}$ momentum tail was previously found for a single-component Fermi gas with odd-parity interactions that are characterized by a scattering length and an effective range~\cite{cui2016high} and for spin-1/2 fermions with both even- and odd-parity zero-range interactions in the presence of a single-particle spin-orbit coupling term~\cite{qin2020universal,wang2024boson}; such a spin-orbit coupled system may be interpreted as corresponding to $\alpha=1/2$ in our framework. 
Our derivation shows that the $k^{-3}$ term is due to an interference 
of two different two-body contributions as well as an interference between a two- and a three-body contribution.

{\em{Conclusion ---}}
We have established an $N$-particle Hamiltonian with two-body zero-range interactions that are characterized by the scattering length
$a_{\text{any}}$, which supports eigenstates with bosonic anyon exchange statistics and fermionic anyon exchange statistics.
Just as bosons and fermions, bosonic anyons and fermionic anyons have different particle exchange statistics.  
The jump discontinuities of the anyonic exchange operator profoundly impact the momentum tail of $N$ identical anyons. Specifically, bosonic anyons and fermionic anyons can be distinguished by their momentum distribution.
 The anyonic momentum tails derived in this work are universal, i.e., they apply to any number of particles
 and systems with and without external confinement.
 Our work generalizes the paradigmatic and  analytically tractable
 1D Bose-Fermi duality to systems with  generalized exchange statistics. This paradigmatic model may help elucidate properties of more conventional anyons, which live in 2D  and are formulated in terms of the braid group~\cite{stern2008anyons}, as well as other non-standard exchange statistics such as para-statistics~\cite{green1953generalized, wang2025particle}, which is formulated 
 in terms of higher-dimensional representations of the permutation group and not tied to a particular space dimensionality.  

{\em{Acknowledgement:}}
We thank Sagarika Basak and Jacob Norris for fruitful discussions. R.H. and T.B. acknowledge support from Okinawa Institute of Science and Technology Graduate University. D.B. acknowledges support by the National Science Foundation through grant number PHY-2409311.

\bibliography{apssamp}

\clearpage

\appendix
\begin{widetext}

\begin{center}
\textbf{\large Supplemental Materials: Universal momentum tail of identical one-dimensional anyons with two-body interactions}

\text{Ra\'ul Hidalgo-Sacoto${}^1$, Thomas Busch${}^1$ and D. Blume${}^{2,3}$}

\textit{${}^1$Quantum Systems Unit, Okinawa Institute of Science and Technology Graduate University, \\ Onna, Okinawa 904-0495, Japan\\
${}^2$Homer L. Dodge Department of Physics and Astronomy, The University of Oklahoma, 440 W. Brooks Street, Norman, Oklahoma 73019, USA\\
${}^3$Center for Quantum Research and Technology, The University of Oklahoma, 440 W. Brooks Street, Norman, Oklahoma 73019, USA}

\end{center}

\section{Properties of sign function and their relevance for tail of momentum distribution}
\label{appendix_sign}

The discontinuities of the anyonic wavefunctions arise from (i) the zero-range boundary conditions and (ii) the gauge phase, which is written in terms of sign functions. In the mathematical literature, the discontinuity of the sign-function $\sign(z_{jl})$ from $z_j-z_l=0^+$ to $z_j-z_l=0^-$ is referred to as a jump discontinuity, step discontinuity, or discontinuity of the first kind~\cite{bleistein1986asymptotic}. The definition of the sign function is given in the main text. The property $\sign^2(z)=1$ for all $z$, including $z=0$, ensures that the probability distributions, which can be measured experimentally, are well defined for all points of the configuration space. The property that $\sign(z)$ is undefined at $z=0$ ensures that the behavior of the anyonic wavefunctions when two particles sit on top of each other is governed by the boundary conditions imposed by the two-body zero-range interactions as opposed to the exchange symmetry condition. The discontinuities of the $\sign$-function need to be treated carefully when calculating the momentum distribution. In preparation for these calculations (see~Appendixes~\ref{appendix_momentumtail}-\ref{appendix_momentumtail2}), we now review select Fourier transform properties of the $\sign$-function.

The Fourier transform of the $\sign$ function is given by the Cauchy principal value (PV) of 
the “conventional” Fourier transform expression~\cite{osgood2019lectures},
\begin{eqnarray}
\label{appendix_eq_sign00}
\text{PV} \left[ \int_{-\infty}^{\infty} dz \sign(z) \exp(- \iu k z) \right] = \frac{-2 \iu}{k}.
\end{eqnarray}
To determine the Fourier transform of $\sign(z) z^n$ for $n=1,2,\cdots$, we insert 
\begin{eqnarray}
z^n \exp(-\iu k z) = \frac{1}{(-\iu)^n} \frac{\partial^n \exp(-\iu k z)}{\partial k^n}  
\end{eqnarray}
into
\begin{eqnarray}
\text{PV} \left[ \int_{-\infty}^{\infty} dz \sign(z) z^n \exp(- \iu k z) \right] .
\end{eqnarray}
Exchanging the order of the integration and differentiation and subsequently using 
Eq.~(\ref{appendix_eq_sign00}), we find, after differentiation, 
\begin{eqnarray}
\label{appendix_eq_sign0}
       \text{PV} \left[  \int^{\infty}_{-\infty} dz \: \mathrm{sign}(z)z^n e^{-\iu k z} \right] = \frac{ 2 n!(-\iu)^{n+1}}{k^{n+1}}.
        \end{eqnarray}

        Equation~(\ref{appendix_eq_sign0}) shows that the Fourier transform of the 
        boundary condition term $\sign(z_{1j})(z_{1j})^n$ scales as $k^{-n-1}$.
        Since the tail of the momentum distribution from two-body discontinuities is obtained by taking the square of the magnitude of 
        the Fourier transform,
       the momentum tail contains products of integrals of the 
        form (\ref{appendix_eq_sign0}), i.e., it contains terms that scale as $k^{-n-n'-2}$.
        This shows that the $k^{-2}$ term of the tail of the momentum distribution is generated by $n=n'=0$, i.e., by the square of the Fourier transform of the
        discontinuous $\sign(z_{1j})$ boundary condition term. Since this 
        is a boundary condition in the two-body odd-parity sector, a $k^{-2}$ tail is characteristic for identical fermions~\cite{cui2016universal}.
        The  $k^{-3}$ term of the tail of the momentum distribution due to  two-body discontinuities is generated by $n=0$ and $n'=1$, i.e., it is generated by “mixing” the 
        $\sign(z_{1j})$ term in the two-body odd-parity sector and the $\sign(z_{1j})z_{1j}=|z_{1j}|$ term in the two-body even-parity sector.
        
The contributions to the momentum tail due to the three-body discontinuities are slightly more challenging to analyze without performing the explicit calculation (see the next section). A key point to note is that the contributions from the three-body discontinuities do not arise from a single Fourier transform but from two subsequent Fourier transforms; the second Fourier transform, which is not immediately visible in the expression for the momentum distribution, arises when performing a coordinate transformation to Jacobi coordinates. With this in mind,  
        there is an additional $k$-dependence in one of the integrals outside of the square of the magnitude in the momentum distribution expression. Alltogether, the result  scales as $k^{-n -n'-2}k^{-m-1 }$, where the 
        terms $k^{-n-1}$ and $k^{-n'-1}$  come from the first 
        Fourier transform, and 
        the term $k^{-m-1}$ 
        from the 
        subsequent Fourier transform. By these arguments (see Secs.~\ref{appendix_momentumtail}-\ref{appendix_momentumtail2} for details), the leading order contribution from three-body discontinuities is $k^{-3}$, which arises when 
        $n=n'=m=0$.

 \section{Detailed derivation of momentum tail of $n_{\alpha,+}(k)$ up to order $k^{-3}$}
 \label{appendix_momentumtail}
 
This  Appendix derives the momentum tail expression $n_{\alpha,+}(k)$ for  bosonic anyons up to order $k^{-3}$ [see Eq.~(\ref{eq_momentumtail_alpha+}) of the main text]. As discussed in the main text, the contributions to the tail of the momentum distribution arise due to two-, three-, and higher-body discontinuities. Specifically, this appendix derives explicit expressions for $T_{\alpha,+}^{(2)}(k)$ and $T_{\alpha,+}^{(3)}(k)$. We also comment on the scaling of momentum tail contributions that arise from higher-body discontinuities.

To evaluate $T^{(2)}_{\alpha, +}(k)$, we
    write out $\Psi_{\alpha,+}(z_1,\cdots,z_N)$
    for the case where $z_2,\cdots,z_N$ are distinct and $z_1$ approaches $z_2$:
    \begin{eqnarray}
    \lim_{z_1 \rightarrow z_2} \Psi_{\alpha,+}(z_1,\cdots,z_N)
    =\underbrace{
\left[    \prod_{l=3}^{N} \hat{S}^{\dagger}_{\alpha/2}(z_{2l}) \right]^2
\left[    \prod_{l=3}^{N-1} \prod_{m>l}^N  \hat{S}^{\dagger}_{\alpha/2}(z_{lm}) \right]
\Psi_+(z_2,z_2,z_3,\cdots,z_N) }_{\mbox{independent of $z_1$}} 
\underbrace{h^{(2)}_{\alpha,+}(z_{12})}_{\mbox{jump discontinuities}}. \:
\end{eqnarray}
The term $h^{(2)}_{\alpha,+}(z_{12}) = \hat{S}_{\alpha/2}^{\dagger}(z_{12}) h^{(2)}_{+}(z_{12})$ arises from the two-body pseudo-potential, which imposes a boundary condition
on the short-distance behavior when $z_1$ and $z_2$ coincide.
With the dependence on $z_1$ isolated, we use
\begin{eqnarray}
\label{eq_product_switch}
\int_{-\infty}^{\infty}dz_1 F(z_1) G(z_1-z_2)  = \int_{-\infty}^{\infty}dz_1 F(z_1+z_2) G(z_1)   
\end{eqnarray}
to rewrite the Fourier transform with regards to the $z_1$ coordinate that appears in $n_{\alpha,+}(k)$:
\begin{eqnarray}
 \int^{\infty}_{-\infty} dz_{1}   e^{-\iu k z_{1}} 
        \hat{{{S}}}^{\dagger}_{\alpha/2}(z_{12}) h^{(2)}_+(z_{12})  =
        e^{-\iu k z_2}  \int^{\infty}_{-\infty} dz_{1}   e^{-\iu k z_{1}} 
        \hat{{{S}}}^{\dagger}_{\alpha/2}(z_{1}) \left(1- \frac{|z_{1}|}{a_{\mathrm{any}}} 
        \right),
                \end{eqnarray}
        where we moved the exponential in $z_2$ outside of the integral over $z_1$. 
        Writing the operator $\hat{S}_{\alpha/2}^{\dagger}(z_1)$
       as
       $\cos(\pi \alpha/2) + \iu \sin( \pi \alpha/2) \sign(z_1)$,
       we see  that the Fourier integral contains two jump discontinuities, namely
       $|z_1|$ and $\sign(z_1)$.
       We thus obtain
       \begin{align}
       \label{eq_help_t2b_1}
       \mbox{PV} \Bigg\{  \int^{\infty}_{-\infty} dz_{1}&   e^{-\iu k z_{1}} 
        \hat{{{S}}}^{\dagger}_{\alpha/2}(z_{1}) \left(1- \frac{|z_{1}|}{a_{\mathrm{any}}} \right) \Bigg\}
        = \nonumber \\
        &\iu \sin \left(\frac{\pi \alpha}{2} \right)  \underbrace{ \mbox{PV} \left\{  \int^{\infty}_{-\infty} dz_{1}   e^{-\iu k z_{1}}  \sign(z_1) \right\}}_{\frac{2}{\iu k}} - \frac{1}{a_{\mathrm{any}}} \cos \left(\frac{\pi \alpha}{2} \right)  
        \underbrace{ \mbox{PV} \left\{  \int^{\infty}_{-\infty} dz_{1}   e^{-\iu k z_{1}} z_1 \sign(z_1) \right\}}_{-\frac{2}{k^2}}.
        \end{align}
        Since $\Psi_{\alpha,+}(z_1,\cdots,z_N)$ factorizes, in the limit $z_1 \rightarrow z_2$, into
        $z_1$-independent and $z_1$-dependent terms and since the jump discontinuities add additively to the tail of the momentum distribution, we can evaluate the square of the magnitude:
       \begin{equation}
       \label{eq_help_t2b_2}
     \left|   \mbox{PV} \left\{  \int^{\infty}_{-\infty} dz_{1}   e^{-\iu k z_{1}} 
        \hat{{{S}}}^{\dagger}_{\alpha/2}(z_{1}) \left(1- \frac{|z_{1}|}{a_{\mathrm{any}}}  \right) \right\} \right|^2
        =  
        4 \sin ^2\left( \frac{\pi \alpha}{2} \right) \frac{1}{k^2} + 8 \sin \left( \frac{\pi \alpha}{2} \right)\cos \left( \frac{\pi \alpha}{2} \right)\frac{1}{a_{\mathrm{any}}k^3}.
        \end{equation}
        Inserting the above ingredients into $T^{(2)}_{\alpha,+}(k)$ and 
       using that $\hat{S}^{\dagger}_{\alpha/2}(z_{lm})\hat{S}_{\alpha/2}(z_{lm})$ is equal to $1$, we find
       \begin{align}
       (N-1)T^{(2)}_{\alpha,+}(k)=
       \Bigg[
       4 \sin^2& \left( \frac{\pi \alpha}{2} \right) \frac{1}{k^2} + 8 \sin \left( \frac{\pi \alpha}{2} \right)\cos \left( \frac{\pi \alpha}{2} \right)\frac{1}{a_{\mathrm{any}}k^3}
       \Bigg]\times\nonumber \\
       &\int_{-\infty}^{\infty}dz_2 \underbrace{N(N-1) \int d z_3 \cdots dz_N |\Psi_+(z_2,z_2,z_3,\cdots,z_N)|^2 }_{g_2(z_2,z_2)}
       + {\cal{O}}(k^{-4}).
       \end{align}
       Recognizing $\int dz g_2(z,z)$ as the two-body contact ${\cal{C}}_2$ of the $N$-body system [see the discussion surrounding Eq.~(\ref{eq_density_capm}) of the main text], our final result
       reads
         \begin{eqnarray}
         \label{eq_tail_T2balpha+}
       (N-1)T^{(2)}_{\alpha,+}(k)=
       \frac{4  {\cal{C}}_2}{k^2}  \sin^2 \left( \frac{\pi \alpha}{2} \right) + \frac{8  {\cal{C}}_2}{a_{\mathrm{any}}k^3} \sin \left( \frac{\pi \alpha}{2} \right)\cos \left( \frac{\pi \alpha}{2} \right)
      +{\cal{O}}(k^{-4}) .
       \end{eqnarray}

To evaluate $T^{(3)}_{\alpha,+}(k)$, we treat---as already noted in the main text---the three-body boundary condition
of $\Psi_{+}(z_1,\cdots,z_N)$ to
leading order. 
In performing the Fourier transform with regards to the coordinate $z_1$ and to subsequently integrate
over the coordinates $z_2$ and $z_3$, we find it useful to 
work with the coordinates $z_{23}=z_2-z_3$ and $Z_{23}=(z_2+z_3)/2$ instead of $z_2$ and $z_3$.
The limit of $z_1$, $z_2$, and $z_3$ approaching each other is taken by first letting $z_1$ approach $Z_{23}$
and subsequently letting $z_{23}$ approach $0$; both limits are taken while $Z_{23}$ is fixed 
and while the coordinates $z_m$
with $m \ge 4$ are distinct from each other and not equal to $z_1$, $z_2$, or $z_3$.
Taking this limit, $\Psi_{+}(z_1,\cdots,z_N)$ goes
to $\Psi_+(Z_{23},Z_{23},Z_{23},z_4,\cdots,z_N)h^{(3)}_+(z_{23},z_{23,1})$,
where $h^{(3)}_+(z_{23},z_{23,1})$ is a purely real function that describes the short-distance behavior in the three-body sector.
In writing the limiting short-distance behavior, we used “internal” Jacobi coordinates
$z_{23}$ and $z_{23,1}$, where $z_{23,1}=Z_{23}-z_1$.

Using $z_{12}=z_1-Z_{23}-z_{23}/2$ and $z_{13}= z_1-Z_{23}+z_{23}/2$, we write
\begin{align}
h^{(3)}_{\alpha, +}(z_{23},z_{23,1}) &=\hat{S}_{\alpha/2}^{\dagger}(z_{12})\hat{S}_{\alpha/2}^{\dagger}(z_{13})\hat{S}_{\alpha/2}^{\dagger}(z_{23})h^{(3)}_+(z_{23},z_{23,1})
= \nonumber \\
&=\hat{S}_{\alpha/2}^{\dagger}(z_{23}) \exp \left[  \frac{\iu \pi \alpha}{2} \sign \left(z_1 -Z_{23}-\frac{1}{2}z_{23} \right) \right]
\exp \left[  \frac{\iu \pi \alpha}{2} \sign \left(z_1 -Z_{23}+\frac{1}{2}z_{23} \right) \right]h^{(3)}_+(z_{23},Z_{23}-z_1).
\end{align}
Since the jump discontinuity of the operator $\hat{S}_{\alpha/2}^{\dagger}(z_{23})$ does not
depend on $z_1$, the Fourier transform over $z_1$ involves only 
$\hat{S}_{\alpha/2}^{\dagger}(z_{12})\hat{S}_{\alpha/2}^{\dagger}(z_{13})h^{(3)}_+(z_{23},Z_{23}-z_1)$. We thus consider
\begin{align}
&\int_{-\infty}^{\infty}dz_1 
\exp(-\iu k z_1) \hat{S}_{\alpha/2}^{\dagger}(z_{12})\hat{S}_{\alpha/2}^{\dagger}(z_{13})h^{(3)}_+(z_{23},z_{23,1})= \nonumber \\
&\int_{-\infty}^{\infty}dz_1 \exp(-\iu k z_1)
\exp \left[  \frac{\iu \pi \alpha}{2} \sign \left(z_1-Z_{23}-\frac{1}{2}z_{23} \right) \right]
\exp \left[  \frac{\iu \pi \alpha}{2} \sign \left(z_1-Z_{23}+\frac{1}{2}z_{23} \right) \right]h^{(3)}_+(z_{23},-(z_1-Z_{23})).\nonumber \\
\end{align}
Using Eq.~(\ref{eq_product_switch}), this can be written as
\begin{align}
\int_{-\infty}^{\infty}&dz_1
\exp(-\iu k z_1) \hat{S}_{\alpha/2}^{\dagger}(z_{12})\hat{S}_{\alpha/2}^{\dagger}(z_{13})h^{(3)}_+(z_{23},z_{23,1}) \nonumber \\
&=\int_{-\infty}^{\infty}dz_1 \exp(-\iu k (z_1+Z_{23}))
\exp \left[  \frac{\iu \pi \alpha}{2} \sign \left(z_1-\frac{1}{2}z_{23} \right) \right]
\exp \left[  \frac{\iu \pi \alpha}{2} \sign \left(z_1+\frac{1}{2}z_{23} \right) \right]h^{(3)}_+(z_{23},-z_1).
\end{align}
The exponential $\exp(-\iu k Z_{23})$ can be readily pulled outside of the integral over $z_1$.
To obtain the contributions to the tail of the momentum distribution,
we need to identify the jump discontinuities of the
function that the Fourier transform is taken of. 
Since 
 $z_1$ cannot simultaneously be equal to $z_{23}/2$ and $-z_{23}/2$ for arbitrary but fixed $z_{23}$, 
 the integral yields two distinct contributions to the momentum
 tail, one when $z_1$ approaches $z_{23}/2$ (in this case, $z_1+z_{23}/2$ is equal to $z_{23}$)
 and another when $z_1$ approaches $-z_{23}/2$ (in this case, $z_1-z_{23}/2$ is equal to $-z_{23}$).
 We thus write
 \begin{align}
\mbox{PV}& \left\{ \int_{-\infty}^{\infty}dz_1 
\exp(-\iu k z_1) \hat{S}_{\alpha/2}^{\dagger}(z_{12})\hat{S}_{\alpha/2}^{\dagger}(z_{13})h^{(3)}_+(z_{23},z_{23,1})
\right\}  \nonumber \\
=&\exp(-\iu k Z_{23})   \exp \left[  \frac{\iu \pi \alpha}{2} \sign \left(z_{23} \right) \right] 
\mbox{PV}  \left\{ \int_{-\infty}^{\infty}dz_1 \exp(-\iu k z_1)
\exp \left[  \frac{\iu \pi \alpha}{2} \sign \left(z_1-\frac{1}{2}z_{23} \right) \right] h^{(3)}_+(z_{23},-z_1)
\right\}
\nonumber \\
&+ \exp(-\iu k Z_{23}) \exp \left[  \frac{\iu \pi \alpha}{2} \sign \left(-z_{23} \right) \right]
\mbox{PV}  \left\{  \int_{-\infty}^{\infty}dz_1 \exp(-\iu k z_1)
\exp \left[  \frac{\iu \pi \alpha}{2} \sign \left(z_1+\frac{1}{2}z_{23} \right) \right] h^{(3)}_+(z_{23},-z_1) \right\}.
\nonumber \\
\end{align}
Using Eq.~(\ref{eq_product_switch}) to rewrite both integrals, we find
\begin{align}
\label{eq_help_threebody}
\mbox{PV}& \left\{ \int_{-\infty}^{\infty}dz_1 
\exp(-\iu k z_1) \hat{S}_{\alpha/2}^{\dagger}(z_{12})\hat{S}_{\alpha/2}^{\dagger}(z_{13})
h^{(3)}_+(z_{23},z_{23,1})
\right\}  \nonumber \\
=& \exp(-\iu k Z_{23})   \exp \left(- \frac{\iu k z_{23}}{2} \right)  \exp \left[  \frac{\iu \pi \alpha}{2} \sign \left(z_{23} \right) \right] 
\underbrace{ \mbox{PV}  \left\{ \int_{-\infty}^{\infty}dz_1 \exp(-\iu k z_1)
\exp \left[  \frac{\iu \pi \alpha}{2} \sign \left(z_1\right) \right]
h^{(3)}_+(z_{23},-z_1-z_{23}/2)
\right\}}_{\iu \sin \left( \frac{\pi \alpha}{2} \right)  h^{(3)}_+(z_{23},-z_{23}/2)\frac{2}{\iu k}= \frac{2}{k} \sin \left( \frac{\pi \alpha}{2} \right)h^{(3)}_+(z_{23},-z_{23}/2) }
 \nonumber \\
&+ \exp(-\iu k Z_{23})  \exp \left( \frac{\iu k z_{23}}{2} \right)  \exp \left[  -\frac{\iu \pi \alpha}{2} \sign \left(z_{23} \right) \right]
\underbrace{\mbox{PV}  \left\{  \int_{-\infty}^{\infty}dz_1 \exp(-\iu k z_1)
\exp \left[  \frac{\iu \pi \alpha}{2} \sign \left(z_1 \right) \right] 
h^{(3)}_+(z_{23},-z_1+z_{23}/2)
\right\}}_{\iu \sin \left( \frac{\pi \alpha}{2} \right) h^{(3)}_+(z_{23},z_{23}/2)\frac{2}{\iu k} = \frac{2}{k} \sin \left( \frac{\pi \alpha}{2} \right) h^{(3)}_+(z_{23},z_{23}/2)}.
\nonumber \\
\end{align}
For determining Cauchy's principal value, we used that neither $h^{(3)}_+(z_{23},-z_1-z_{23}/2)$ nor
$h^{(3)}_+(z_{23},-z_1+z_{23}/2)$  have a
jump discontinuity at lowest order as $z_1$ approaches zero.
We now take the absolute value square of the right hand side of Eq.~(\ref{eq_help_threebody}) and 
subsequently integrate over $z_{23}$. In writing the result, we use that $\hat{S}_{\alpha/2}(z_{23}) \hat{S}_{\alpha/2}^{\dagger}(z_{23})=1$ and keep only terms that contribute to the power law decay of the tail of the momentum distribution:
\begin{align}
\label{eq_help_threebody3}
\mbox{PV}& \left\{ 
\int_{-\infty}^{\infty} dz_{23}   \left| 
 \int_{-\infty}^{\infty}dz_1 
\exp(-\iu k z_1) \hat{S}_{\alpha/2}^{\dagger}(z_{12})\hat{S}_{\alpha/2}^{\dagger}(z_{13})h^{(3)}_+(z_{23},z_{23,1})
\right|^2 \right\}
 \nonumber \\
&=\frac{4}{k^2} \sin^2 \left( \frac{\pi \alpha}{2} \right) 
\left[\mbox{PV} \left\{
\int_{-\infty}^{\infty} dz_{23} \exp \left(- \iu k z_{23} \right) \exp \left[ \iu \pi \alpha \sign(z_{23})  \right] h^{(3)}_+(z_{23},z_{23}/2)h^{(3)}_+(z_{23},-z_{23}/2)
\right\} + \: \mathrm{c.c.} \right].
\end{align}

To determine which terms contribute power law decays to the tail, we need to analyze the behavior
of  $h^{(3)}_+(z_{23},z_{23}/2)h^{(3)}_+(z_{23},-z_{23}/2)$ as $z_{23}$ approaches $0$. 
We know that the full wave function must be symmetric under the exchange of any pair of particles.
From this, it follows that neither $h^{(3)}_+(z_{23},z_{23}/2)$ nor $h^{(3)}_+(z_{23},-z_{23}/2)$
possesses a jump discontinuity at leading order.
Setting $h^{(3)}_+(z_{23},z_{23}/2)=h^{(3)}_+(z_{23},-z_{23}/2)=1$, 
 we find up to order $k^{-3}$:
\begin{align}
\mbox{PV}& \left\{ 
\int_{-\infty}^{\infty} dz_{23}   \left| 
 \int_{-\infty}^{\infty}dz_1 
\exp(-\iu k z_1) \hat{S}_{\alpha/2}^{\dagger}(z_{12})\hat{S}_{\alpha/2}^{\dagger}(z_{13})h^{(3)}_+(z_{23},z_{23,1})
\right|^2 \right\}
\nonumber \\
&=\frac{4}{k^2} \sin^2 \left( \frac{\pi \alpha}{2} \right) 
\bigg[\underbrace{\mbox{PV} \left\{
\int_{-\infty}^{\infty} dz_{23} \exp \left(- \iu k z_{23} \right) \exp \left[ \iu \pi \alpha \sign(z_{23})  \right]
\right\}}_{\iu \sin \left( \pi \alpha \right) \frac{2}{\iu k}=\frac{2}{k} \sin(\pi \alpha)} + \: \mathrm{c.c.}\bigg]=
\frac{16}{k^3} \sin^2 \left( \frac{\pi \alpha}{2} \right) \sin(\pi \alpha).
\end{align}
We are now ready to evaluate $T^{(3)}_{\alpha,+}(k)$:
\begin{align}
\frac{(N-1)(N-2)}{2} T^{(3)}_{\alpha,+}(k)=
\frac{16}{k^3} \sin ^2&\left( \frac{\pi \alpha}{2} \right) \sin(\pi \alpha)\times\nonumber \\
&\int_{-\infty}^{\infty} dZ_{23} \underbrace{ N \frac{(N-1)(N-2)}{2} \int dz_4 \cdots dz_N 
|\Psi_{+}(Z_{23},Z_{23},Z_{23},z_4,\cdots,z_N)|^2}_{\frac{1}{2}g_3(Z_{23},Z_{23},Z_{23})}.
\end{align}
Recognizing $\int dz g_3(z,z,z)$ as the three-body contact ${\cal{C}}_3$ of the $N$-body system, we 
find
\begin{eqnarray}
\label{eq_tail_T3balpha+}
\frac{(N-1)(N-2)}{2} T^{(3)}_{\alpha,+}(k)
=
\frac{8 {\cal{C}}_3}{k^3} \sin ^2\left( \frac{\pi \alpha}{2} \right) \sin(\pi \alpha) + {\cal{O}}(k^{-4}).
\end{eqnarray}

Adding up the two body and three body discontinuity contributions up to order $k^{-3}$, we have
        \begin{eqnarray}
        \label{eq_final_bosonicanyon}
       \lim_{|k| \rightarrow \infty} n_{\alpha,+}(k) =
       \frac{4  {\cal{C}}_2}{k^2}  \sin^2 \left( \frac{\pi \alpha}{2} \right) + 
       \frac{4  {\cal{C}}_2}{a_{\mathrm{any}}k^3}
         \sin \left( \pi \alpha\right)
        +  \frac{8 {\cal{C}}_3}{k^3} \sin ^2\left( \frac{\pi \alpha}{2} \right) \sin(\pi \alpha)  
      +{\cal{O}}(k^{-4}) .
       \end{eqnarray}
This result is reported in Eq.~(\ref{eq_momentumtail_alpha+}) of the main text.

The derivation above accounts for contributions to the tail of $n_{\alpha,+}(k)$ up to order $k^{-3}$. The derivation shows that the contributions to the tail due to two-body and three-body discontinuities are associated with the two-body Tan contact ${\cal{C}}_2$ and the three-body Tan contact ${\cal{C}}_3$, respectively. Assuming that the short-distance behavior is fully governed by the two-body scattering length (i.e., that effects associated with the two-body effective range and three- and higher-body scattering parameters can be neglected), we can draw additional conclusions about the scaling of the tail of anyonic momentum distributions. Using that the units of $n_{\alpha,+}(k)$ are length and those of the $M$-body Tan contact ${\cal{C}}_M$ are length$^{-M+1}$, we deduce that the leading-order contribution to the momentum tail due to $M$-body discontinuities is proportional to $k^{-M}$. Using that $a_{\text{any}}$ is the only length scale in the problem, sub-leading contributions scale as $(a_{\text{any}}k)^{-1}k^{-M}$. Table~\ref{tab_scaling} summarizes the scaling.

\begin{table}[t]
        \centering
        \begin{tabular}{l|l|l}
             & Tan contact  & momentum tail contributions \\ \hline
$T^{({2})}_{\alpha, \pm}$ & $\mathcal{C}_2$ (units of length$^{-1}$) & $\propto k^{-2}$, $\propto a_{\text{any}}^{-1}k^{-3}$, $\propto a_{\text{any}}^{-2}k^{-4}$, $\propto a_{\text{any}}^{-3}k^{-5}$, $\dots$ \\
            $T^{({3})}_{\alpha, \pm}$ & $\mathcal{C}_3$ (units of length$^{-2}$) & $\propto k^{-3}$, $\propto a_{\text{any}}^{-1}k^{-4}$, $\propto a_{\text{any}}^{-2}k^{-5}$,  $\dots$ \\
            $T^{(4)}_{\alpha, \pm}$ & $\mathcal{C}_4$ (units of length$^{-3}$) & $\propto k^{-4}$, $\propto a_{\text{any}}^{-1}k^{-5}$, $\dots$ \\
            $T^{(5)}_{\alpha, \pm}$  & ${\cal{C}}_5$  (units of length$^{-4}$) & $\propto k^{-5}$, $\dots$
        \end{tabular}
        \caption{Contributions to the tail of the momentum distribution of bosonic anyons and fermionic anyons due to two-, three-, four-, and five-body discontinuities. The proportionality factors, which are not reported,  depend on the statistical factor $\alpha$.}
        \label{tab_scaling}
    \end{table}

   \section{Derivation of momentum tail up to order $k^{-3}$}
   \label{appendix_momentumtail2}

To determine the tail of $n_{\alpha,-}(k)$, we need to replace $\Psi_-$ in the previous section by $\Psi_+$.
To write $\Psi_-(z_1,\cdots,z_N)$, we take advantage of the Bose-Fermi mapping, i.e., we  write
 $\Psi_-(z_1,\cdots,z_N)=\hat{A}(z_1,\cdots,z_N)\Psi_{+}(z_1,\cdots,z_N)$.
Next, we use that $\hat{A}$ can be expressed in terms of $\hat{{\cal{S}}}_{1/2}^{\dagger}$, 
\begin{eqnarray}
\hat{A}(z_1, \dots, z_N) = (-i)^{N(N-1)/2}\hat{\mathcal{S}}^\dagger_{1/2}(z_1, \dots, z_N).
\end{eqnarray}
It then follows
\begin{eqnarray}
\label{eq_anyon_combine}
    \Psi_{\alpha,-}(z_1, \dots, z_N) = (-i)^{N(N-1)/2}\hat{\mathcal{S}}^\dagger_{(\alpha + 1)/2}(z_1, \dots, z_N)\Psi_{+}(z_1, \dots, z_N).
    \end{eqnarray}
    We emphasize that $\alpha$ takes, as before, values from $0$ to $1$. Equation~(\ref{eq_anyon_combine}) tells us that, mathematically, the fermionic anyon wavefunctions can be constructed by applying the $\hat{\cal{S}}^{\dagger}$ operator with argument $(\alpha+1)/2$ to $\Psi_+$.
    As a consequence, the tail of $n_{\alpha,-}(k)$ can be obtained by changing  $\alpha/2$ in Eq.~(\ref{eq_final_bosonicanyon}) to $(\alpha+1)/2$.
    The result is
        \begin{eqnarray}
        \label{eq_final_fermionicanyon}
       \lim_{|k| \rightarrow \infty} n_{\alpha,-}(k) =
       \frac{4  {\cal{C}}_2}{k^2}  \cos^2 \left( \frac{\pi \alpha}{2} \right) - 
       \frac{4  {\cal{C}}_2}{a_{\mathrm{any}}k^3}
         \sin \left( \pi \alpha \right)
       -  \frac{8 {\cal{C}}_3}{k^3} \cos ^2\left( \frac{\pi \alpha}{2} \right) \sin(\pi \alpha)  
      +{\cal{O}}(k^{-4}). 
       \end{eqnarray}
This result is reported in Eq.~(\ref{eq_momentumtail_alpha-}) of the main text. We note that Eq.~(\ref{eq_final_fermionicanyon}) can alternatively be derived by following the approach employed in Sec.~\ref{appendix_momentumtail}. Adapting the derivation of Sec.~\ref{appendix_momentumtail} as appropriate, we carried out this brute-force approach and arrived, as we should, at the same result.

\section{Validation of general framework through two examples} \label{app_validation}

To confirm our predictions for the anyonic momentum tails, we calculate the full momentum distribution for two examples and compare their asymptotic large-$|k|$ behaviors with 
Eqs.~(\ref{eq_final_bosonicanyon}) and (\ref{eq_final_fermionicanyon}).

\subsection{Two bound anyons in free space}

 As a first example, 
we calculate the momentum distribution corresponding to the bound state of two identical anyons in free space with zero-range interactions ($a_{\mathrm{any}}>0$)  and vanishing center-of-mass momentum.
Since the center-of-mass momentum is assumed to be zero, the center-of-mass wave function
contributes merely a normalization factor of $1/\sqrt{L}$, where $L$ 
 denotes a length that will be 
taken to infinity at the end of the calculation. We denote the relative wave function of two identical bosons  that interact through the zero-range pseudopotential with scattering length $a_+$  by $\psi^{(2b)}_{+}(z_{12})$, 
\begin{eqnarray}
    \psi^{(2b)}_{+}(z_{12}) = \frac{1}{\sqrt{a_{+}}} \exp \left( - \frac{|z_{12}|}{a_+} \right).
\end{eqnarray}
Constructing the  bound state wave function of two identical bosonic anyons by applying ${\cal{N}}_{\alpha}\hat{{{S}}}_{\alpha/2}^{\dagger}(z_{12})$  to  $\psi^{(2b)}_{+}(z_{12})$,
renaming $a_+$ by $a_{\text{any}}$, 
and evaluating Eq.~(\ref{eq:momentum_correlationfunction1}) from the main text, we find 
  \begin{eqnarray}
\label{eq:twoboundanyonsmomentumdistribution_freespace}
    n^{(2b)}_{\alpha,+}(k) = 8a_{\mathrm{any}}\frac{\left[\cos(\frac{\pi \alpha}{2} ) + a_{\mathrm{any}}k\sin(\frac{\pi \alpha}{2})\right]^2}{(1+a_{\mathrm{any}}^2k^2)^2}.
\end{eqnarray}
To construct the bound state of two identical fermionic anyons, we start with  $\psi^{(2b)}_{\alpha,-}(z_{12}) = \sign(z_{12})\psi^{(2b)}_{\alpha,+}(z_{12})$. A straightforward calculation then yields
\begin{eqnarray}
\label{eq:twoboundanyonsmomentumdistribution_freespace2}
    n^{(2b)}_{\alpha,-}(k) = 8a_{\mathrm{any}}\frac{\left[\sin(\frac{\pi \alpha}{2} ) - a_{\mathrm{any}}k\cos(\frac{\pi \alpha}{2})\right]^2}{(1+a_{\mathrm{any}}^2k^2)^2}.
\end{eqnarray}

Expanding 
Eqs.~(\ref{eq:twoboundanyonsmomentumdistribution_freespace}) and
(\ref{eq:twoboundanyonsmomentumdistribution_freespace2})
for large $|k|$, we find that the large momentum tail up to order $k^{-3}$ agrees with Eqs.~(\ref{eq_final_bosonicanyon}) and (\ref{eq_final_fermionicanyon}), respectively.
To arrive at this conclusion, we use that
the two-body contact ${\cal{C}}_2$ of the bound state of two particles interacting with a zero-range pseudopotential with scattering length $a_{\text{any}}$ is~\cite{barth2011tan}
\begin{eqnarray}
    {\cal{C}}_2= \frac{2}{a_{\mathrm{any}}}.
\end{eqnarray}
In addition, we set ${\cal{C}}_3$ in Eqs.~(\ref{eq_final_bosonicanyon}) and (\ref{eq_final_fermionicanyon}) to zero.

\subsection{Three bound anyons in free space}

To explicitly confirm that the three-body Tan contact ${\cal{C}}_3$ enters into $n_{\alpha,\pm}(k)$ at order $k^{-3}$, we calculate the full momentum distribution of the bound state of three identical anyons 
in free space for vanishing center-of-mass momentum.   

The relative bound-state wave function
$\psi^{(3b)}_+(z_{12},z_{13},z_{23})$ of three identical bosons with zero-range interactions ($a_{\text{any}}>0$) reads~\cite{mcguire1964study}
\begin{eqnarray}
    \psi^{(3b)}_+(z_{12},z_{13},z_{23})=
    \sqrt{\frac{8}{3 a_{+}^2}} \exp \left( - \frac{|z_{12}|}{a_{+}}- \frac{|z_{13}|}{a_+}- \frac{|z_{23}|}{a_+} \right).
\end{eqnarray}
The two- and three-body contacts 
of the three-body wave function $L^{-1/2}\psi^{(3b)}_{\alpha,+}$
are
\begin{eqnarray}
    {\cal{C}}_2=\frac{8}{a_{\mathrm{any}}}
\end{eqnarray}
and 
\begin{eqnarray}
    {\cal{C}}_3=\frac{16}{a_{\mathrm{any}}^2};
\end{eqnarray}
here, we used again that the Tan contact is independent of the exchange statistics. The momentum  
distributions
of $\psi^{(3b)}_{\alpha, \pm}$ 
are
computed 
by 
considering 
each of the six 
permutation sectors of $z_1$, $z_2$, and $z_3$ separately
and taking the $L\rightarrow \infty$ limit at the end of the calculation. For 
bosonic anyons,
we find
\begin{align}
\label{eq_tail_3b+}
    n^{(3b)}_{\alpha, +}(k) =& \frac{16 a_{\mathrm{any}}}{(4 + 
   a_{\mathrm{any}}^2 k^2 )^2 (16 + a_{\mathrm{any}}^2 k^2 )} \times \nonumber \\
   &\bigg[(8 - a_{\mathrm{any}}^2 k^2 ) (4 + a_{\mathrm{any}}^2 k^2 ) \cos\left(\pi \alpha\right) - 
   16 (-2 + a_{\mathrm{any}}^2 k^2 ) \cos\left(2 \pi \alpha\right)  \nonumber \\
   &\quad+(4 + a_{\mathrm{any}}^2 k^2 ) (10 + a_{\mathrm{any}}^2 k^2  + 
      6 a_{\mathrm{any}} k \sin\left(\pi \alpha \right)) - 
   2 a_{\mathrm{any}} k (-20 + a_{\mathrm{any}}^2 k^2 ) \sin\left(2 \pi \alpha\right)\bigg].
\end{align}
For fermionic anyons, in contrast, we find, 
\begin{align}
\label{eq_tail_3b-}
    n^{(3b)}_{\alpha, -}(k) =& \frac{16 a_{\mathrm{any}}}{(4 + 
   a_{\mathrm{any}}^2 k^2 )^2 (16 + a_{\mathrm{any}}^2 k^2 )}\times \nonumber \\
   &\bigg[(-8 + a_{\mathrm{any}}^2 k^2 ) (4 + a_{\mathrm{any}}^2 k^2 ) \cos\left(\pi \alpha\right) - 
   16 (-2 + a_{\mathrm{any}}^2 k^2 ) \cos\left(2 \pi \alpha\right)  \nonumber \\
   &\quad +(4 + a_{\mathrm{any}}^2 k^2 ) (10 + a_{\mathrm{any}}^2 k^2 - 
      6 a_{\mathrm{any}} k \sin\left(\pi \alpha \right)) - 
   2 a_{\mathrm{any}} k (-20 + a_{\mathrm{any}}^2 k^2 ) \sin\left(2 \pi \alpha\right)\bigg].
\end{align}
The anyonic momentum distributions $n^{(3b)}_{\alpha, \pm}(k)$ are plotted in Fig.~\ref{fig:anyon-map}(b) of the main text [the figure and surrounding discussion suppresses the superscript $(3b)$ for notational convenience].

Taylor expanding $n_{\alpha,\pm}^{(3b)}(k)$, Eqs.~(\ref{eq_tail_3b+}) and (\ref{eq_tail_3b-}), we find that the large momentum tails up to order $k^{-3}$ agree with Eqs.~(\ref{eq_final_bosonicanyon}) and (\ref{eq_final_fermionicanyon}). 
Figure~\ref{app_fig:momentum tail} analyzes the positive-$k$ tail of $n^{(3b)}_{\alpha,+}(k)$. The top right and bottom right panels show $a_{\text{any}} k^2 n^{(3b)}_{\alpha,+}(k)$ and $a^2_{\text{any}} k^3 n^{(3b)}_{\alpha,+}(k)$, respectively, for $\alpha=0$. The fact that the scaled momentum distributions for $\alpha=0$, i.e., for two identical bosons, approach zero for large $k$ shows that $k^{-2}$ and $k^{-3}$ tails are absent. This is consistent with the results of Ref.~\cite{olshanii2003short}. The upper left and lower left panels show $n^{(3b)}_{\alpha,+}(k)$, using various scalings, for $\alpha=1/2$ (anyons) and $\alpha=1$ (fermions). The upper left panel shows that the calculated momentum distributions (solid lines) follow the predicted asymptotic $k^{-2}$ behavior (dashed lines). For convenience we introduced the abbreviations
\begin{eqnarray}
\label{eq_c2_abbreviaton}
    C_2(\alpha) = \mathcal{C}_2 \sin^2(\pi \alpha/2)
\end{eqnarray}
and 
\begin{eqnarray}
\label{eq_c3_abbreviation}
    C_{2,3}(\alpha) = \mathcal{C}_2 \sin(\pi \alpha)/a_\mathrm{any} + 2\mathcal{C}_3 \sin^2(\pi \alpha /2)\sin(\pi \alpha).
\end{eqnarray} 
The lower left panel shows that the bosonic anyon momentum tail possesses a $k^{-3}$ tail that follows the analytically derived behavior (dashed line) and that the fermionic system ($\alpha=1$) does not possess a $k^{-3}$ contribution. 
\begin{figure}[tb]
    \centering
    \includegraphics[width=0.95\linewidth]{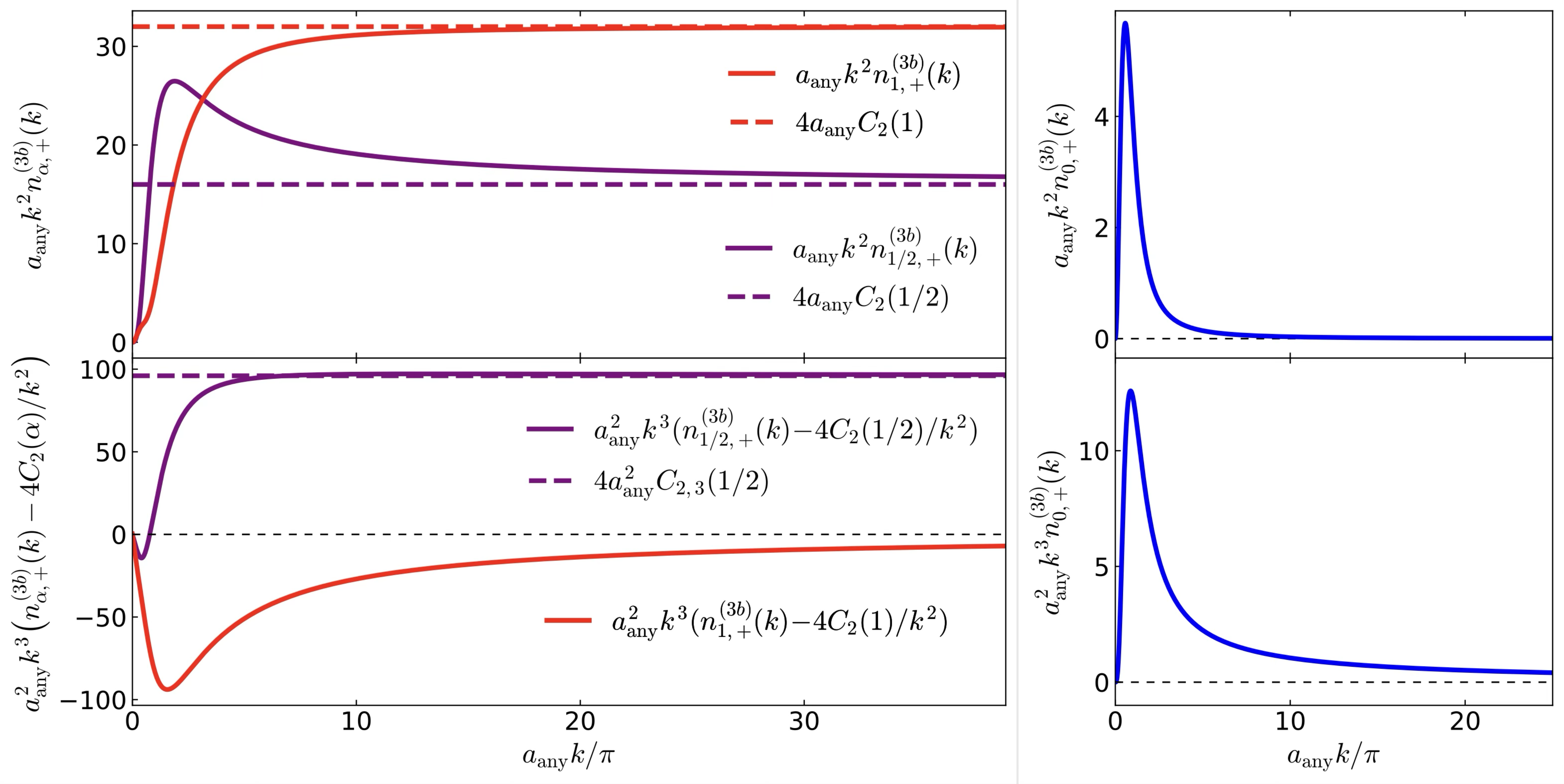}
    \caption{Analysis of momentum distribution $n_{\alpha,+}^{(3b)}(k)$ for three bound bosonic anyons with scattering length $a_{\text{any}}$. Solid lines show  $n_{\alpha,+}(k)$ [Eq.~(\ref{eq_tail_3b+}); scaled and, in some cases, shifted] for various $\alpha$ ($\alpha=1/2$ and $1$ in the left column and $\alpha=0$ in the right column) while dashed lines show the analytically derived asymptotic tails [Eq.~(\ref{eq_final_bosonicanyon})]. The coefficients $C_2(\alpha)$ and $C_3(\alpha)$ are defined in Eqs.~(\ref{eq_c2_abbreviaton}) and (\ref{eq_c3_abbreviation}), respectively. At large $k$, excellent agreement is found between the full (scaled and shifted) momentum distribution and the analytically derived tail. 
Top row: To analyze the leading-order contribution to the tail, the momentum distribution is multiplied by $k^2$. For $\alpha=1$ (upper left), bosonic anyons approach their dual complement, namely “ordinary” fermions; in this case, the $k^{-2}$ term contributes to the tail. For $\alpha=0$ (upper right), bosonic anyons reduce to “ordinary” bosons; correspondingly, the $k^{-2}$ contribution is absent. 
Bottom row: To analyze the sub-leading-order contribution to the tail, the leading-order tail contribution  is subtracted and the resulting expression is subsequently multiplied by $k^3$. For $\alpha=1/2$, the $k^{-3}$ term contributes to the tail. For $\alpha=1$ and $\alpha=0$, the $k^{-3}$ contribution is absent.}
    \label{app_fig:momentum tail}
\end{figure}

\newpage

 \end{widetext}

\end{document}